# Anomalous dispersion in pulsar 21-cm radiation reveals the existence of faster-than-c phenomena in near field of scatterers

Mark E. Perel'man*)

*Racah Institute of Physics, Hebrew University, Jerusalem, Israel*

**ABSTRACT**

At passage of 21-cm pulsar radiation through clouds of neutral hydrogen atoms the signal in the region of anomalous dispersion is appearing as faster-than-c. As unlike laboratory researches separate scatterers are located on big distances from each other, this effect can be attributed only to the consecutive independent scattering on isolated atoms. For its explanation we must accept that photons are emitted and absorbed on the distances λ/2, in a near field, instantaneously. Such peculiarity of near field has been established earlier within the frame of QED and explains, quantitatively and qualitatively, different "superluminal" observations. It shows that processes of absorption and reemission of photons do not submit to requirements of special relativity describing only uniform movements, and consequently possibilities of faster-than-c phenomena in the near field are not excluded.

**INTRODUCTION**. During two last decades several contra-intuitive observations of light pulses propagation faster than light in vacuum have been revealed. As was usually presented, these observations contradict the main postulate of relativity (paradox of "superluminality"), and many attempts of their coordination with the axioms of theory were undertaken (e.g. [1] and references therein).

However the conclusions on the base of these observations about possible infringements of relativity requirements are, generally speaking, very exaggerated.

First, let us remind that the special relativity describes only and only the uniform movement. Therefore the emission (and absorption) of particles, and also wave transitions from one environment into another, i.e. almost all problems connected with near fields, can be not describable, at least completely, by the special relativity. Notice that all uncountable checks had been carried out usually in far fields, on such distances when a contribution of near fields is negligible.

*). *E-mail:* m.e.perelman@gmail.com.

Secondly, let's consider the process of single photon emission (absorption must be similarly described). In accordance with the quantum paradigm, the emission process cannot be fulfilled as a gradual release of single photon's energy by a source, since in such case would be possible to interfere with a course of this process. Hence the photon should be radiated entirely, i.e. instantaneously at least on the distance λ/2, necessary and sufficient for process ending: this distance can be considered as the border of near field, i.e. as the effective size of scatterer. Thus, the emission must be carried out as the quantum jump, i.e. as "the nonlocality in the small", the strictly spatially limited within the near field [2, 3]. But even so must be underlined that this process should be sufficiently far from the uncertainty limits, i.e. it would be represented as an independent measurable phenomenon.

Thus, it becomes impossible to speak about photon's speed in a near field, and by taking into account the length of this jump the quantitative description of all observable superluminal phenomena becomes possible [2, 3]. Notice that such jumps would be summed in the phenomenon of the frustrated total internal reflection (FTIR) that must lead to even bigger superluminal pulse advancing [4] (cf. [5]).

Natural complexity in the interpretation of all early experiments is that they were fulfilled at photons propagation through environments where besides effect of superluminal distribution can play role a tunneling, pulse reformation, etc. The most demonstrative manifestation of the reality of this phenomenon would be the observation of a superluminal pulse at light passage through a rarefied cloud of separate elementary scatterers in the absence of tunneling effects and so on.

Exactly such unique observation in the region of anomalous dispersion of 21-cm pulsar radiation at passage through HI regions of neutral hydrogen atoms is described in [6]. This discovery requires a returning to the faster-than-c problem on the more definite base and represents the main aim of the letter.

It is needed to note that all other observations that seem contra-intuitive are executed also in the regions of anomalous dispersion or at observation of light transitions from one medium into another, including the phenomena of FTIR (in all cases close to singularities or jumps of optical dispersion). Thus it is possible to assume that all these unusual observations are related solely to the features of near field.

As these "superluminal" phenomena are described (with the analyses of early observations) in our cited articles, we shall begin with a brief description of some features of temporal magnitudes needed for the examined problem. Then the kinetic approach to light pulse propagation will be described that determines an arrival time in dependence on frequency and with taking into account the features of elementary acts, based on [7, 8]. It allows a comparison of the estimated and observed data. Some general problems are mentioned in the Conclusions.

**DURATION OF ELEMENTARY SCATTERING ACTS**. A transparent way to introduction of durations concept for examined problem is such (comp. [9]). For the case of uniform stationary linear media the sufficiently weak incoming and outgoing signals are connected by the integral convolution:

$$O(t,\mathbf{r}) = \int dt' d\mathbf{r}' \; R(t - t', \mathbf{r} - \mathbf{r}') \; I(t', \mathbf{r}'). \qquad (1)$$

The decomposition of logarithm of response function $R(\omega, \mathbf{r})$ into series near to $\omega_0$ leads to the appearance of temporal functions:

$$\partial \ln R / i\partial\omega = \tau(\omega, \mathbf{r}) \equiv \tau_1 + i\,\tau_2. \qquad (2)$$

Restoration of response function and substitution of its inverse Fourier transformation in (1) shows that $\tau_1$ is the time-delay during elastic scattering (e.g. [9]) and $\tau_2$ is the duration of final state formation [8]. The physical significance of $\tau_2$ becomes more transparent at its formulation as $\tau_2 = \partial \ln|R|/\partial\omega$: hence, this magnitude can be considered as a measure of temporary incompleteness of the final (of the free photon's) state.

Let us begin with the simplest oscillator model of response function, e.g. with the dielectric susceptibility:

$$R(\omega, \mathbf{r}) = A(\mathbf{r})/[(\omega_0 - i\gamma/2)^2 - \omega^2], \qquad (3)$$

where $\omega = 2\pi f = 2\pi c/\lambda$ is the angular frequency, $\gamma$ is the line-width. This form seems adequate for almost classical description of rarefied gas of neutral hydrogen atoms in the region of unique spin-flip frequency $\omega_0$.

Both temporal functions can be represented by (2) as

$$\tau_1(\omega) \simeq \gamma/2[(\omega_0 - \omega)^2 + \gamma^2/4]; \qquad (4)$$

$$\tau_2(\omega) \simeq (\omega - \omega_0)/[(\omega_0 - \omega)^2 + \gamma^2/4]. \qquad (5)$$

At $\gamma \to 0$ these functions have the limiting values:

$$\tau_1(\omega) \to \pi\delta(\omega - \omega_0), \qquad \tau_2(\omega) \to 1/(\omega - \omega_0). \qquad (6)$$

So $\tau_1$ shows that the delay at elastic scattering does not change parameters of photon. The function $\tau_2$ is positive for the case of normal dispersion and negative in the anomalous dispersion region, its absolute value is twice bigger the uncertainty limit and therefore must be measurable.

The in-depth analysis of temporal functions must be executed in the frame of quantum electrodynamics [3, 8]. The free pass of photon is describable by the causal propagator of QED represented as $D_c(t, \mathbf{r}) = \overline{D} + iD_1$, where the first Green function is supported in the light cone, but $D_1$ oversteps the limits of cone and hence represents the prime interest for us. (The propagator $\overline{D}$ corresponds to the classical relativistic theory and $D_1$ represents quantum additions to it.) In the mixed representation $D_1(\omega, \mathbf{r}) = \frac{\operatorname{sgn}\omega}{2\pi}\sin(\omega r)$ and corresponding temporal function at $\omega > 0$

$$\tau(\omega, \boldsymbol{r}) = (\partial/i\partial\omega)\ln D_1(\omega, \boldsymbol{r}) = -i\,(r/c)\cot(\omega r/c), \tag{7}$$

or

$$\tau_1(\omega, \boldsymbol{r}) = 0, \qquad \tau_2(\omega, \boldsymbol{r}) = -(r/c)\cot(\omega r/c). \tag{8}$$

These expressions implicitly show absence of delays outside of cone. At transition $\omega \to \omega - \omega_0 \equiv \delta\omega$ and with the expansion of cotangent

$$\tau_2(\omega, \boldsymbol{r}) = -\frac{1}{\delta\omega} - 2\sum_1^\infty \frac{\delta\omega}{(\delta\omega)^2 - n^2(\pi c/r)^2}. \tag{9}$$

It shows the existence of poles beyond the resonance, with $\delta\omega \neq 0$. The first of them is on the distance $r_1 = \pi c/\delta\omega$ corresponding to $\Delta\ell = \lambda/2$. As $\tau_2$ can be negative, this process can be instantaneous; it corresponds to the jump of "photon" at the act of emission (absorption) by free electron, along the photon formation path. (Notice that a simple substitution of $\omega \to \omega + i\gamma/2$ into (7) leads in the first order to expressions similar to (4) and (5)).

For processes on atomic electrons complete duration includes equal durations of the state formation at absorption and emission and the time delay on scatterer:

$$\Delta\tau = 2\tau_2 + \tau_1 \simeq \frac{2\delta\omega + \gamma/2}{\delta\omega^2 + \gamma^2/4}\,. \tag{10}$$

This expression shows that the phenomenon of advancing, i.e. the jump of photon, must be executed in the restricted part of region of anomalous dispersion: $\omega < \omega_0$, $|\delta\omega| < \gamma/4$.

**KINETICS OF OPTICAL DISPERSION**. The classical approach to phenomena of an optical dispersion is based on scattering of a falling wave on all scatterers of medium and the subsequent interference of all secondary waves [10]. Such presentation is developed at description of medium as enough dense formation, in which distance between scatterers is less than wave length and nothing hinders to an interference of secondary waves.

If, however, a medium is so rarefied that these distances is much bigger than wave length, the statistical approach becomes doubtful and the usage of kinetic, microscopic consideration based on the quantum scattering theory seems preferable. Such approach to the phenomena of an optical dispersion has been offered in the article [7], in more details it is described in [8, 11].

In the microscopic approach is accepted that the photon flies by the free path length $\ell$ with vacuum speed $c$, stays on a scatterer for a certain time $\tau$ and then continues its flight. The length of free flight is defined (if all scatterers are of the one type of density $\rho$) as

$\ell(\omega) = 1/\rho\sigma,$ (11)

where for scattering on free electrons the classical Thompson cross-section $\sigma_T$ can be taken. For the resonance scattering of photon onto neutral atom the cross-section of resonance fluorescence (e.g. [12]) can be taken:

$$\sigma_{res} \simeq \sigma_T \frac{\omega^2/4}{(\omega-\omega_0)^2+\gamma^2/4}\,. \qquad (12)$$

If the complete time of delay of a single photon of frequency ω on an everyone scatterer is $\tau(\omega)$, the time of passage of the distance $L$ is equal to

$$T(\omega) = T_0 + \Delta T; \quad T_0 = l/c; \quad \Delta T = (L/\ell)\tau = L\rho\sigma\tau. \qquad (13)$$

This estimation immediately leads to the group velocity of light:

$$u = \frac{L}{T(\omega)} = \frac{c}{1 + c\rho\sigma\tau}\,, \qquad (14)$$

i.e. to the group index of refraction $n_{gr} = c/u$ and to the usual index of refraction:

$$n_{gr}(\omega) = \frac{d}{d\omega}\omega n(\omega) \quad \text{or} \quad \nu(!) = \frac{1}{\omega}\int \text{?}\gamma\rho(!)d\omega \qquad (15)$$

with the natural condition νγρ(1) = 1. At conditions of normal dispersion $n_{gr} \geq n$, but in the anomalous dispersion region $n_{gr} \leq n$, which, in particular, may be conditioning by $\tau < 0$.

**<u>Observable advancing</u>**. The deducted refraction index allows calculating of the velocity of light passing through clouds of HI, in dependence on its frequency just as in [6]. However the performed considerations allow direct estimation of the advancing in the region of anomalous dispersion without a reference to refraction indices.

The simplest expression for complete durations of transition through HI clouds of summary depth $L$

$$\Delta T(\omega) \simeq L\rho\sigma_T \frac{[2(\omega-\omega_0)+\gamma/2]\,\omega^2/4}{[(\omega-\omega_0)^2+\gamma^2/4]^2}. \qquad (16)$$

There are, of course, several capabilities of its specification. As the atoms are moving with respect to the observer, $\omega_0$ would be replaced by $\omega_c = \omega_0(1 - v_c/c)$, its Doppler-shifted value, and $\Delta T$ must be averaged over $\delta\omega = \omega - \omega_0$ with taking into account the temperature of gas, the mean free path in (1) must be, in general, specified as $\ell \rightarrow \ell' = \ell + 2c|\tau_2|$, and so on. But all these corrections are small enough and at the analysis of principal effect can be omitted.

Moreover, as we are especially interested in the range $\delta\omega = \omega - \omega_0$, $\gamma/4 < |\delta\omega| < \gamma$, let's take $|\delta\omega| = \gamma/2$ (other values do not essentially change its order):

$\Delta T(\omega) \sim \frac{3}{2} L\rho\, \sigma_T\, \omega^2/\gamma^3.$ (17)

For $\omega = 2\pi\cdot 1440$ MHz, $\gamma \sim (10^{-5} \div 10^{-6})\,\omega$ and the observed advancing $\Delta T \leq 20$ microsecond brings to $L\rho \leq 2\cdot(10^{13} \div 10^{10})\text{cm}^{-2}$. If the density of neutral H atoms $\rho \sim 1 \div 10^4$ cm$^{-3}$, it allows to estimate the limits of neutral clouds dimensions that seems non impossible.

For the scattering on free electrons in these clouds $\tau \sim 1/\omega$ and

$\Delta T_{elect}\,(\omega) \sim L\rho_{electr}\ \sigma_T/\omega.$ (19)

The absence of frequency selectivity and smallness of this magnitude complicates its direct measurement.

**CONCLUSIONS**. Thus such principal conclusion can be formulated: *The elastic scattering of photon (its absorption and reemission) in the definite part of the region of anomalous dispersion, where the momentum of virtual photon surpasses its energy,, is executable via instantaneous jumps onto λ/2 at absorption and at reemission in the scope of scatterer's near field.*

These results can be considered as corresponding to the more general theorem, established in [2]: *Superluminal transfer of excitations through a linear passive substance can be affected by nothing but by the instantaneous tunneling of virtual particles; the tunneling distance is of order of half a wavelength corresponding to the deficiency in the energy relative to the nearest stable (resonance) state. The nonlocality of the electromagnetic field must be described by the 4-potential* $A_\mu$, *whereas the fields* **E** *and* **B** *fields remain unconnected to the near field.* In the examined case it requires only an evident reformulation: excitations can be replaced by formatted photons, etc.

We stress that our description corresponds to the Wigner's formulation of causality: *"The scattered wave cannot escape a scatterer before the initial wave reaches it"* [13] since the "scatterer" must undeniably include its own near field of the order of λ. It means that the effective sizes of scatterer depend on the scattering frequency and on its correspondence to the inner structure of scatterer. This formulation is optimally suiting the quantum measurement paradigm and seems more adequate than the conventional one: the standard point wise causality formulation contradicts quantum theory that does not admit such strict localization of emitting or absorbing points.

It is necessary to emphasize that the introduction of "nonlocality in the small", within the limits of a near field zone, is not a priori unacceptable in the framework of QED. Indeed the principle of locality was verified experimentally only in the far field zone, for **E** and **B**. Hence, the assumption of a possible nonlocality of parts of the electromagnetic field, not included in its (transverse) far field, is not evidently forbidden.

Besides of it can be noted that the described "nonlocality in the small" can be contained in the condition of gauge invariance: the classical Lorentz condition, $\partial A_\mu/\partial x_\mu = 0$, is replaced, in QED, by the Lorentz-Fermi condition $\partial A_\mu/\partial x_\mu |0\rangle = 0$ that requires the vanishing of the "superfluous" components of $A_\mu$, the "pseudophotons", *only on the average*. Hence, it does not exclude the possibility of nonlocality of superfluous parts of the field in the near zone (it is proved in [3]).

Described phenomena must exist at scattering of other particles also. Notice that this effect, in particular, must be taken into account at consideration of the light propagation through gravitational fields also (e.g. [14]).

These phenomena can be considered via high temporal derivatives of propagators. Temporal functions corresponding to odd derivatives, including velocities and jerks, lead to jumps, etc. (the most evidently needs of higher derivatives are presented in gravity, e.g. [15]). But we omit here their considerations.

How can be interpreted the results of this and previous articles on the maximal speed of interactions and the relativistic causality?

It can be stated that the macroscopic speed approaches asymptotically and very rapidly to *ç* from above by diminishing the role of the near field zone or, more correctly, by separating from it. Seemingly, the exceeding of *ç* in the near field zone of a source, just as tunneling processes, can not be described by the classical relativity: the postulate of relativity in its classical form remains completely correct in the area of its applicability, namely for far fields and, generally speaking, outside near fields and tunneling areas, i.e. outside the regions of non-uniform movements.

Thus, the so-called "paradox of superluminality" is completely resolved.

## REFERENCIES

1. H. Winfil. Phys. Rep., **436**, 1(2006); N. Borjemscaia et al. Opt. Express, **18**, 2279 (2010).
2. M.E. Perel'man, Ann. Phys. (Leipzig), **14**, 733 (2005); Int. J. Theor. Phys., **46**, 1277 (2007).
3. M.E. Perel'man, Ann. Phys. (Leipzig), **16**, 350 (2007).
4. M.E. Perel'man, Phys. Lett. A, **373**, 648 (2009).
5. N. Brunner, V. Scarani, M. Wegmüller, M. Legré and N. Gisin, Phys. Rev. Lett., **93**, 203902 (2004); D. Salart, A. Baas, C. Branciard, N. Gisin and H. Zbinden, Nature, **454**, 861 (2008).
6. F. A. Jenet, D. Fleckenstein, A. Ford, A. Garcia, R. Miller, J. Rivera and K. Stovall. Astrophys. J., **710**, 1718 (2010), In: arXiv/ 0909.2445v3.
7. M.E. Perel'man and G.M. Rubinstein, Sov. Phys. Dokl., **203**, 798 (1972).
8. M.E. Perel'man, Int. J. Theor. Phys., **47**, 468 (2008).
9. M.L. Goldberger and K.M. Watson, Collision Theory. Wiley, New York (1964).
10. M. Born and E. Wolf, Principles of Optics, Pergamon, New York, 1959.

[11]. M.E. Perel'man, Quantum Kinetics: Duration of Basic Elementary Processes. Nova Sc. Publ. – in press.

[12]. W. Heitler, The Quantum Theory of Radiation.1954.

[13]. E.P. Wigner, Phys. Rev., **98**, 145 (1955).

[14]. A. A. Abdo et al. Nature, **462**, 331 (2009).

[15]. M. Visser. Class. Quantum Grav., **21**, 2603 (2004).